# Performance Analysis of Optimized VANET Protocols in Real World Tests

Jamal Toutouh, Enrique Alba
Dept. de Lenguajes y Ciencias de la Computació´n
University of Ma´laga, ETSI Informa´tica
Campus de Teatinos, Ma´laga - 29071, Spain
{jamal, eat}@lcc.uma.es

# Performance Analysis of Optimized VANET Protocols in Real World Tests

Jamal Toutouh, Enrique Alba
Dept. de Lenguajes y Ciencias de la Computació´n
University of Ma´laga, ETSI Informa´tica
Campus de Teatinos, Ma´laga - 29071, Spain
{jamal, eat}@lcc.uma.es

*Abstract*—Vehicular *ad hoc* networks (VANETs) provide the communications required to deploy *Intelligent Transportation Systems* (ITS). In the current state of the art in this field there is a lack of studies on real outdoor experiments to validate the new VANETs protocols and applications proposed by designers. In this work we have addressed the definition of a testbed in order to study the performance of the *Vehicular Data Transfer Protocol* (VDTP) in a real urban VANET. The VDTP protocol has been tested by employing six different parameter settings: one defined by human experts and five automatically optimized by means of metaheuristic algorithms (PSO, DE, GA, ES, and SA). As a result, we have been able to confirm the performance improvements when optimized VDTP configurations are used, validating the results previously obtained through simulation.

## I. INTRODUCTION

Integrating the capabilities of new generation wireless technologies to vehicles and roadside infrastructure is the main purpose of *vehicular ad hoc networks* (VANETs). The idea is to provide *Vehicle-to-Vehicle* (V2V) and *Vehicle-to-Infrastructure* (V2I) communications that enable *Intelligent Transportation Systems* (ITS) [1]. These systems represent an opportunity to develop powerful applications such as cooperative traffic monitoring, prevention of collisions, control of traffic flows, blind crossing, and real-time detour routes computation, among others [2]. Other applications based on multimedia content distribution can be defined, such as downloading movie trailers from nearby cinemas, virtual tours of hotel rooms, etc. Such applications are very demanding and require optimized services of data exchange among the nodes.

Recent researches focus their effort on the design and specification of information dissemination strategies for VANETs, usually in the form of different unicasting, multicasting, and broadcasting approaches [3], [4]. In this context, the *Vehicular Data Transfer Protocol* (VDTP) [5] arose from the work done on the *CARLINK EUREKA-CELTIC European Project* (http://carlink.lcc.uma.es). VDTP is an application layer protocol that allows the *end-to-end* file transfer specifically designed for VANETs.

In VANETs, the changes in the weather and the traffic conditions, the motion of the vehicles, and the presence of different types of obstacles affect the propagation of the signals. Additionally, this generates frequent topology changes and network fragmentation. These factors constitute the main challenges that the network designers have to face. In order to address these problems, it is crucial to be equipped with efficient communication protocols, previously optimized in indoor *CARLAB* experiments by simulation.

Obtaining optimal protocol configurations by using exact techniques is practically impossible due to the enormous number of possible configurations. Therefore, the use of automatic intelligent tools is required to face these problems. Metaheuristic algorithms [6] emerged as efficient techniques able to tackle complex optimization problems. Indeed, these algorithms present successful performance dealing with multitude of engineering problems [7], [8]. Unfortunately, the use of metaheuristics in VANETs is still limited, and just a few a related approaches can be found in the literature.

In Garc´ıa-Nieto et al. [9], five metaheuristic algorithms were employed to find the optimal configuration of VDTP file transfer protocol in realistic VANET scenarios. More recently, Toutouh et al. [10] applied also a metaheuristic technique (*Differential Evolution*) to find efficient parameters settings of the *Optimized Link State Routing* (OLSR) protocol in order to adapt this routing protocol for use in urban VANETs.

Currently, in the VANETs specialised literature, the evaluation of the VANET's protocols and applications is carried out mainly by means of simulations [3], [11], [12]. There is a lack of studies using outdoor experiments that could validate them and obtain more significant results like bandwidths, coverage, data loss, protocols overload, etc. depending on the real situations. This is because it is difficult due to several issues concerning available resources and accurate performance analysis. Indeed, it is neither easy nor cheap to have a high number of real vehicles and a real scenario for only practical purposes [13].

In the present work, we are aimed at defining a testbed in order to study the performance of the VDTP file transfer service between cars in a real urban VANET. In these outdoor experiments, the VDTP protocol has been tested following different parameter settings: the optimal VDTP configurations proposed in Garc´ıa-Nieto et al. [9] and the standard one proposed by CARLINK experts [5]. Thus, the results offer the possibility to confirm the quality-of-service (QoS) improvements on a VANET's performance when optimized protocols are used, validating the results obtained through simulations.

The main contributions of this work are:

- Defining and deploying a real world urban VANET.
- Developing a file transfer software application that follows the VDTP protocol definition.
- Studying the performance of the VDTP file transfer service when exchanging several file types in different VANET situations.
- Using different metrics and statistical tools to compare the VDTP QoS when the different proposed configurations are used.

The remaining of this paper is organised as follows. In the next section, we include the details about the testbed definition and the experimental settings. Section III presents the results, performance analysis, and comparisons achieved after the experimentation. Finally, conclusions and future work are drawn in Section IV.

## II. TESTBED OVERVIEW

Exchanging up-to-date traffic information and multimedia contents are two salient applications of VANETs. This requires sending and receiving data files of different type and length. Thus, in the present work, we analyse the QoS of different VDTP configurations transferring several types of data files. Thus, we have defined a set of outdoor *CARLAB* experiments performed in a real urban VANET scenario.

### A. VANET Scenario Definition

The *ad hoc* network deployed in our testbed is comprised of two nodes (cars) moving through some selected roads in an area of 1,440,000 $m^2$ from the downtown of Má laga (Spain). During the experimentation, the drivers of each of mobile node followed the path marked by the dotted lines in Fig. 1. The roads were opened to the general traffic during a non rush hour, so the number of vehicles travelling through our scenario were not constant (see Fig. 2). In turn, they were regulated by several traffic lights and signals. Therefore, the speed and the distance between the vehicles varied over time, just as it would be in any city. According to the tracking information, during the experiments the average distance between the nodes was 77 meters.

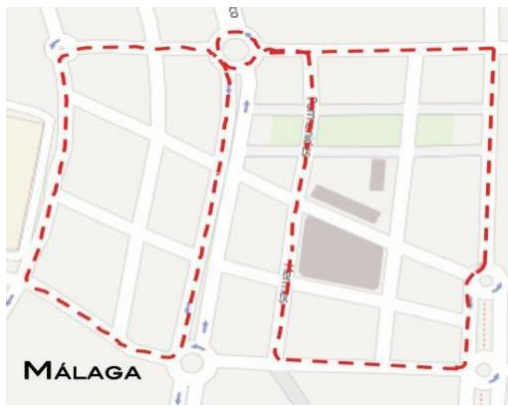

Fig. 1. Area in Má laga where the experiments were carried out.

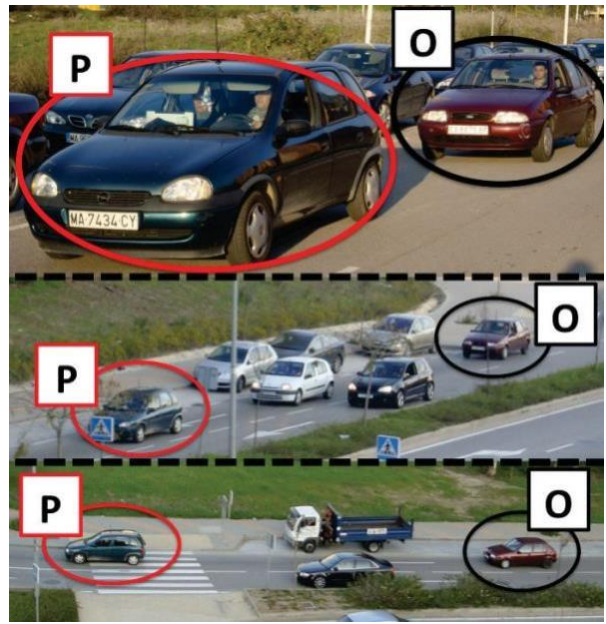

Fig. 2. Nodes during the experiments (*P=petitioner* and *O=owner*).

In our outdoor *CARLAB* experiments, the nodes were cars carrying a notebook equipped with a PROXIM ORiNOCO PCMCIA (IEEE 802.11bg) WiFi transceiver (http://www.proxim.com) connected to a 7 dBi omni-directional antenna located on roof top of the car (see Fig. 3). The global network were configured following the VANET specifications used in the simulations presented in García-Nieto et al. [9] (see Table I). The file transfers were performed by using an application developed in Java that follows the VDTP protocol definition. Additionally, we included a GPS unit in each vehicle to track their movement.

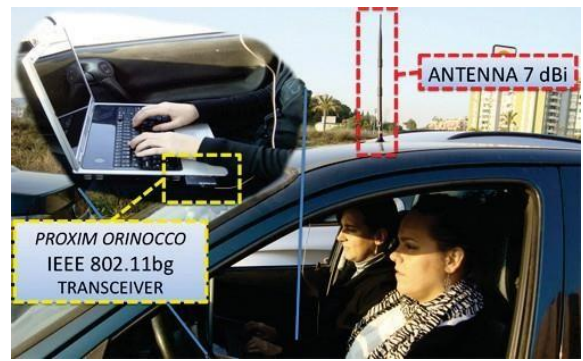

Fig. 3. Vehicles equipment used to perform the file transfers.

TABLE I
VANET INSTANCE SPECIFICATION

| Parameter | Value |
| --- | --- |
| Propagation model | Two Ray Ground |
| Carrier Frequency | 2.4 GHz |
| Channel bandwidth | 5.5 Mbps |
| MAC Protocol | IEEE 802.11b |
| Routing Protocol | DSR |
| Transport Protocol | UDP |
| Application Protocol | VDTP |

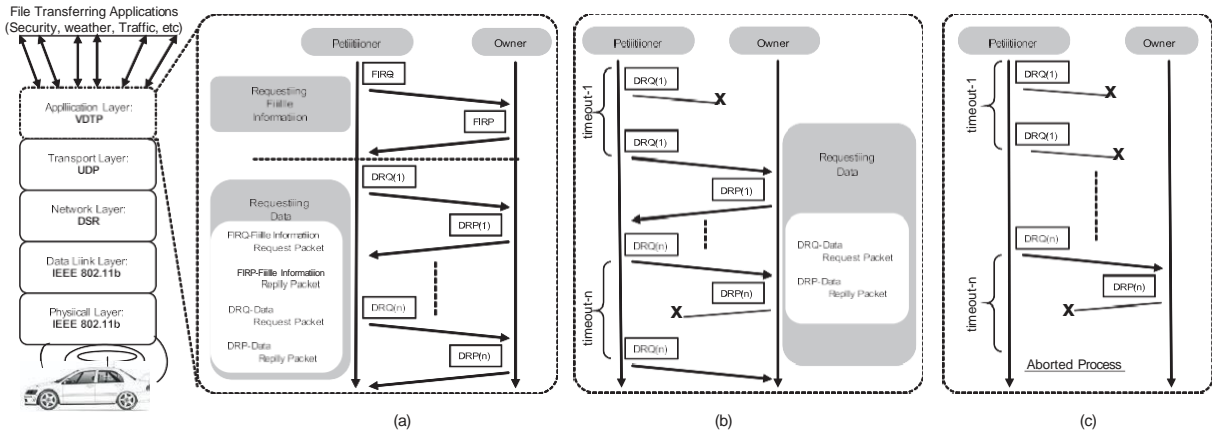

Fig. 4. VDTP operation modes: (a) a complete file exchange is done; (b) timeout expiration and retransmission; (c) communication refused.

## B. Vehicular Data Transfer Protocol

VDTP [5] is an application layer protocol that allows the *end-to-end* file transfer to be used in VANETs. It operates on a transport layer protocol, typically UDP. This implies that considerations about the interconnection modes and routing issues can be avoided, since they are carried out by the previous down layer protocols.

In VDTP, the communication process is carried out by both, the file requester node (**petitioner**), which tries to download a file, and the file **owner** node, which stores the file (see Fig. 4). This transfer protocol operates by using the following packets: *FIRQ (File Information Request)*, *FIRP (File Information Reply)*, *DRQ (Data Request)*, and *DRP (Data Reply)*.

Once the file petitioner knows the name and the location of a given file, it starts the communication by using the FIRQ packet in order to obtain the file metadata. Then, the petitioner waits for this metadata which is provided by the file owner through the FIRP packet. After receiving the information about the file, the petitioner node starts the transfer by sending the DRQ(1) packet to request the first file segment (chunk). Then, it waits for the first chunk sent by the owner by means of DRP(1) packet. This operation is repeated by both, petitioner and owner, until transferring the last chunk DRP($n$), and hence making up the complete file (see Fig. 4.a).

The VANETs communications are carried out in a hostile environment which can provoke packet loss during the communication process. In order to solve such issues, VDTP provides with mechanisms based on timers and counters. The *timeout* mechanism controls the waiting time until resending a concrete DRQ or FIRQ (see Fig. 4.b). The *counter* mechanism controls the maximum number of retransmissions of a concrete DRQ/FIRQ. The file transference between the vehicles is refused when this maximum number is reached (see Fig. 4.c).

This protocol is configured by means of three parameters:

- *chunk_size*: The size of the segments in which the files are split to be sent by the file owner node.
- *retransmission_time*: The *timeout* time waited by the petitioner node before resending FIRQ or DRQ.
- *max_attempts*: The maximum *counter* value.

## C. Experimental Settings

The experiments carried out in this work are aimed at studying the performance transferring files by using different VDTP configurations in a real VANET. Specifically, we have tested the six different VDTP configurations presented in Table II. Five of these configurations were obtained automatically in [9] by using five different metaheuristic algorithms: *Particle Swarm Optimization* (PSO), *Differential Evolution* (DE), *Genetic Algorithm* (GA), *Evolutionary Strategy* (ES), and *Simulated Annealing* (SA). The *EXPERTS* configuration was proposed by the experts in the scope of *CARLINK EUREKA-CELTIC European Project*.

TABLE II
VDTP PARAMETERS SETTINGS TAKEN INTO ACCOUNT IN THIS STUDY.

| Configuration Name | VDTP Configuration Parameters | | |
|---|---|---|---|
| | chunk_size (Bytes) | retransmission_time (seconds) | max_attempts |
| **PSO** | 41,358 | 10.00 | 3 |
| **DE** | 28,278 | 6.00 | 9 |
| **ES** | 23,433 | 10.00 | 8 |
| **GA** | 31,196 | 3.83 | 9 |
| **SA** | 19,756 | 6.43 | 3 |
| **EXPERTS** | 25,600 | 8.00 | 8 |

In order to perform a general study, we have taken into account five types of data files of different sizes: 100 KBytes and 500 KBytes typical in traffic information services; and 1 MByte, 5 MBytes, and 10 MBytes that contain multimedia content. For each VDTP configuration, the vehicles exchange 15 files of each type, i.e., for each VDTP configuration there are 75 (5×15) file transfers.

In turn, we have defined two different types of experiments named *Urban Low Speed tests* and *Urban High Speed tests* to study the influence of the speed in the performance of the VANET. In the first ones, the vehicles speed fluctuates between 20 km/h and 30 km/h. In the *Urban High Speed tests*, the vehicles speed fluctuates between 40 km/h and 50 km/h.

After the file transfers, we have measured the global information about the VDTP QoS in terms of the *number of lost packets* during the downloads, the *transmission time* required, and the *amount of data* exchanged between vehicles. This last two metrics are employed to compute the *effective transmission data rate* of the network during the file transfers.

TABLE III
AVERAGE NUMBER OF LOST PACKETS PER TRANSFER AND EFFECTIVE TRANSMISSION DATA RATES ACHIEVED BY THE DIFFERENT VDTP CONFIGURATIONS DURING THE EXPERIMENTS

| File Size | VDTP Configuration | Urban Low Speed (20-30 km/h) | | Urban High Speed (40-50 km/h) | |
|---|---|---|---|---|---|
| | | Lost Packets (Data) | Effective Data Rate (KBytes/s) | Lost Packets | Effective Data Rate (KBytes/s) |
| 100 KBytes | PSO | 0.0 | 409.071 | 0.0 | 398.852 |
| | DE | 0.0 | 436.008 | 0.0 | **476.131** |
| | ES | 0.0 | 423.882 | 0.0 | 423.320 |
| | GA | 0.1 | 445.342 | 0.0 | 414.768 |
| | SA | 0.0 | **493.717** | 0.0 | 402.369 |
| | EXPERTS | 0.0 | 456.934 | 0.0 | 429.721 |
| | AVERAGE | 0.016 | 444.159 | 0.0 | 424.193 |
| 500 KBytes | PSO | 0.0 | 631.454 | 0.0 | 654.180 |
| | DE | 0.0 | **697.256** | 0.0 | 645.269 |
| | ES | 0.2 | 470.885 | 0.0 | 651.626 |
| | GA | 0.0 | 677.265 | 0.0 | 677.705 |
| | SA | 0.0 | 646.538 | 0.0 | **693.088** |
| | EXPERTS | 0.0 | 664.802 | 0.0 | 641.828 |
| | AVERAGE | 0.033 | 631.367 | 0.0 | 660.616 |
| 1 MByte | PSO | 0.0 | 717.596 | 0.0 | 696.155 |
| | DE | 0.2 | 582.832 | 0.1 | 628.457 |
| | ES | 0.0 | **723.764** | 0.0 | 684.709 |
| | GA | 0.2 | 679.841 | 0.0 | **715.286** |
| | SA | 0.0 | 704.318 | 0.1 | 708.165 |
| | EXPERTS | 0.0 | 691.338 | 0.0 | 678.908 |
| | AVERAGE | 0.066 | 683.281 | 0.033 | 676.232 |
| 5 MBytes | PSO | 0.0 | 698.371 | 0.2 | **659.464** |
| | DE | 0.0 | **740.345** | 0.1 | 623.026 |
| | ES | 0.2 | 668.905 | 0.2 | 643.788 |
| | GA | 0.0 | 688.538 | 0.3 | 620.872 |
| | SA | 0.2 | 563.724 | 0.2 | 656.115 |
| | EXPERTS | 0.2 | 600.207 | 0.2 | 574.677 |
| | AVERAGE | 0.1 | 660.015 | 0.2 | 629.657 |
| 10 MBytes | PSO | 0.4 | 643.576 | 0.4 | **621.259** |
| | DE | 0.4 | **690.145** | 0.4 | 609.239 |
| | ES | 0.2 | 618.988 | 0.5 | 604.702 |
| | GA | 0.6 | 650.997 | 0.6 | 613.021 |
| | SA | 0.6 | 578.297 | 0.7 | 581.800 |
| | EXPERTS | 0.5 | 606.760 | 0.6 | 532.147 |
| | AVERAGE | 0.45 | 631.460 | 0.533 | 593.695 |
| *GLOBAL* | | 0.133 | 610.056 | 0.153 | 598.878 |

## III. RESULTS

This section presents the *CARLAB* experimental results from two points of view. First, we study the communications carried out between the cars during the experiments in order to evaluate the VDTP service. Next, we discuss about the performance of each VDTP parameters settings taken into account in this work in order to compare them with each other.

### A. VANET Global Performance

After the experimentation, we observed that all files were transferred completely and correctly. Mainly this was because the two cars (network nodes) were always following the same course with possible cars between them (see Figure 2). Therefore, even though there were lost packets because of networks problems related with the distance or the existence of obstacles between the nodes, they were able to reconnect with each other before the file transfers were refused.

Table III presents all the results obtained during the whole experimentation. It shows the average number of lost packets during the transference of a file and the average effective transmission data rate performed during the downloads. The results are grouped by the VDTP parameterisation used (PSO, DE, ES, GA, SA, and EXPERTS), the file type transferred (100 KBytes, 500 KBytes, 1 MByte, 5 MBytes, and 10 MBytes), and both experiment types.

Globally, in terms of transmission data rates, the majority of the file transfers were carried out with a competitive bandwidth higher than 600 KBytes/s. In addition, we can check that the communications performed better (larger data rates and smaller packet loss) when the speeds of the vehicles were lower. In Table III (last row), we can observe that, on average, during *Urban Low Speed tests* there were 0.133 lost packets per file transfer with an effective data rate of 610.056 KBytes/s. In contrast, during *Urban High Speed tests* there were more packet loss (0.1533) and lower bandwidth (598.878 KBytes/s).

Additionally, we have studied the influence of the file size in the effective transmission data rates of the downloads. On average, the 1 MByte files are transferred with the higher bandwidth for the two experiment types (see Fig. 5). These files were transferred with an effective data rate of 683.281 KBytes/s during the *Urban Low Speed tests* and 676.232 KBytes/s during the *Urban High Speed tests* (see Table III). This is because the VDTP protocol configurations used in this study were optimized to exchange data stored in 1 MByte files [9]. The transfers of files of 500 KBytes and 5 MBytes were exchanged with data rates between 629.657 KBytes/s and 660.616 KBytes/s.

The smallest data rates were obtained when the 100 KBytes files were transferred. On average, they were 444.159 KBytes/s during the *Urban Low Speed tests* and 424.193 KBytes/s during the *Urban High Speed tests*. Transferring this type of file using the PSO configuration and when the vehicles were moving at high speed was achieved the lowest bandwidth (398.852 KBytes/s). This is because of the impact of the handshaking process of VDTP (FIRP and FIRQ packets exchange), that is greater for the smaller files. The lesson learned is that, as the nodes speed, the sizes of the files to transfer determine the performance of the VANET.

*B. VDTP Configuration Comparison*

In this section we compare the performance of the different VDTP configurations taken into account in this study: the experts one and those computed automatically by means of metaheuristic algorithms in García-Nieto et al. [9]. The comparison has been made in terms of VDTP effective transmission data rates or *goodput*. In addition, we have studied the data loss suffered during the experiments.

Fig. 5 shows the *goodput* achieved during the experiments transferring files using VDTP protocol configured by experts (EXPERTS) and employing the parameter settings obtained by using metaheuristics, for low and high speeds. However, checking the information provided by this figure and by Table III, it is not evident how to provide any global conclusion about which is the configuration that performed the best.

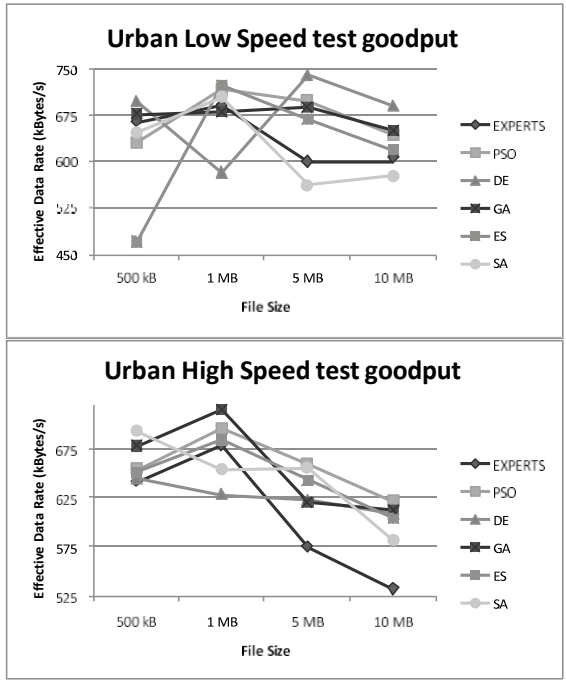

Fig. 5. Effective transmission data rates (goodput) (KBytes/s) achieved during the experiments transferring files using VDTP protocol configured by human experts (EXPERTS) and metaheuristics (PSO, DE, GA, ES, and SA), for both, low and high speeds (without taking into account transfers of 100 KB files).

As we have a distribution of the effective transmission data rates of each VDTP configuration (75 goodput values for each VDTP configuration) for each test, we have applied statistical tests in order to provide a comparison with statistical confidence. Therefore, we have applied the Friedman Ranking statistical test [14] to the *Urban Low Speed test* goodput results, to the *Urban High Speed test* goodput results, and to the goodput results without taking into account the nodes speed. Thus, we have been able to compare the VDTP configurations in different contexts.

We have used this non-parametric test since the resulting distributions violated the conditions of equality of variances (heteroscedasticity) several times and thus ANOVA is not appropriate [14]. The confidence level was set to 95% ($p\text{-}value$=0.05), which allows us to ensure that all these distributions are statistically different if they result in $p\text{-}value<0.05$.

TABLE IV
FRIEDMAN RANK TEST WITH CONFIDENCE LEVEL 95%

| Urban Low Speed Configuration | Rank | Urban High Speed Configuration | Rank | Urban Configuration | Rank |
|---|---|---|---|---|---|
| **PSO** | **4.26** | **PSO** | **4.26** | **PSO** | **4.26** |
| GA | 3.95 | SA | 3.62 | ES | 3.60 |
| DE | 3.73 | ES | 3.56 | GA | 3.54 |
| ES | 3.64 | DE | 3.28 | DE | 3.51 |
| SA | 2.74 | EXPERTS | 3.16 | SA | 3.18 |
| EXPERTS | 2.68 | GA | 3.12 | EXPERTS | 2.92 |

The results of Friedman Ranking test (see Table IV) ranked the VDTP parameterization returned by PSO as the best one during both, low and high speed tests. PSO is also the configuration with higher data rates. The average goodput of this configuration was 620.014 KBytes/s for *Urban Low Speed tests*, 605.982 KBytes/s for *Urban High Speed tests*, and 612.998 KBytes/s without taking into account the nodes speed.

EXPERTS configuration of VDTP showed the worst rank among the compared VDTP configurations for the transmission data rates for *Urban Low Speed tests*, achieving a goodput of 604.008 KBytes/s. In turn, this configuration presented the second worst rank for *Urban High Speed tests* with a goodput of 571.456 KBytes/s. Globally, the experts configuration showed the worst rank (see Table IV) and the lowest average goodput (587.732 KBytes/s).

This confirms the results obtained by means of simulations presented in García-Nieto et al. [9] that presented the PSO configuration as the best one. In turn, CARLINK experts configuration achieved the lowest bandwidth during the simulations carried out in this article.

In terms of data loss, ES configuration lost the lowest amount of data and GA suffered with the highest data loss for all experiments (see Table V). This is related with the trade off between chunk size and retransmission time configuration parameters. On the one hand, ES configuration employs the longest retransmission time (10 seconds) and the second smallest chunk size (23, 433 bytes). On the other hand, GA configuration defines a retransmission time of 3.83 seconds (the shortest one) and the second largest chunk size (31, 196 bytes).

TABLE V
TOTAL LOST DATA (IN BYTES) DURING THE EXPERIMENTS.

| Urban Low Speed | | Urban High Speed | | Sum of All Lost Data | |
|---|---|---|---|---|---|
| Configuration | Lost data | Configuration | Lost data | Configuration | Lost data |
| ES | 14,060 | ES | 16,403 | ES | 30,462 |
| SA | 15,804 | DE | 16,967 | DE | 33,933 |
| PSO | 16,543 | SA | 19,756 | SA | 35,560 |
| DE | 16,967 | EXPERTS | 20,480 | EXPERTS | 38,400 |
| EXPERTS | 17,920 | PSO | 24,815 | PSO | 41,358 |
| GA | 28,076 | GA | 28,076 | GA | 56,152 |

PSO is the second worst VDTP configuration in terms of data loss. However, this has not effect on the global performance of VDTP transfers because, on average, PSO exchanged the data files correctly and completely taking the shortest transmission times and, therefore, presenting the highest effective data rates (goodput).

IV. CONCLUSION

In this work we have addressed the definition of a testbed to study the performance of the VDTP file transfer protocol in a real urban VANET. For this task, we have analysed the file exchange of different file types carried out by two cars moving through open roads in downtown of Ma´laga (Spain). We have studied the communication performances through two different types of experiments taken into account the speed of the nodes (*Urban Low Speed tests* and *Urban High Speed tests*).

In addition, we have compared six proposed VDTP configurations, one defined by the experts of *CARLINK EUREKA-CELTIC European Project* and five optimized ones by means of metaheuristics (PSO, DE, GA, ES, and SA). In the light of the experimental results we can conclude that:

- The VDTP protocol is presented as a competitive solution for transferring files between vehicles in a VANET.
- The VDTP based software application was able to transfer all the files completely and correctly, carrying out the transfers with an average goodput of $603.842$ KBytes/s.
- The use of counters and timers of VDTP allows the transfers were complete hiding the possible problems caused by packet loss.
- As expected, the communications performed better (larger data rates and smaller packet loss) when the speeds of the network nodes (cars) were lower.
- 1 MByte files were exchanged achieving the highest bandwidth, being $683.281$ KBytes/s during *Urban Low Speed tests* and $676.232$ KBytes/s during *Urban High Speed tests*.
- Analysing the QoS obtained for each VDTP parameter settings applied in our experiments, we have observed that globally the automatically optimized VDTP configurations performed better than the one proposed by the human experts. These results were confirmed applying the Friedman Rank statistical test to the network goodput, that ranked the VDTP parameters settings returned by PSO as the best one. In turn, for any urban speed, the human experts configuration showed the worst rank.
- Finally, the results obtained by means of performing the outdoor *CARLAB* experiments (real VANET communications) confirm the ones obtained by the indoor *CARLAB* experiments (employing VANET simulations). Thus, VANET simulation is presented as an accurate tool to evaluate protocols and applications of this kind of networks to be combined with real outdoor tests.

Thus, the metaheuristic techniques seem to be adequate to address the problem of configuring VANET protocols, because they automatically obtain efficient parameter settings.

As a matter of the further work, we are currently extending our testbed by increasing the number of vehicles connected that belong to the urban VANET scenario presented here, i.e., adding multi-hop communications. Also, we are planning the definition of a highway VANET scenario in order to test the file transfers in this kind of roads.

ACKNOWLEDGMENT

Authors acknowledge funds from the Spanish Ministry MICINN and FEDER under contract TIN2008-06491-C04-01 (M* http://mstar.lcc.uma.es) and CICE, Junta de Andaluc´ıa, under contract P07-TIC-03044 (DIRICOM http://diricom.lcc.uma.es).